# A Hamiltonian Approach to Floating Barrier Option Pricing

**Qi Chen[1] · Hong-tao Wang[2] · Chao Guo[2]**

## Abstract

Hamiltonian approach in quantum mechanics provides a new thinking for barrier option pricing. For proportional floating barrier step options, the option price changing process is similar to the one dimensional trapezoid potential barrier scattering problem in quantum mechanics; for floating double-barrier step options, the option price changing process is analogous to a particle moving in a finite symmetric square potential well. Using Hamiltonian methodology, the analytical expressions of pricing kernel and option price could be derived. Numerical results of option price as a function of underlying price, floating rate, interest rate and exercise price are shown, which are consistent with the results given by mathematical calculations.

## 1 Introduction

In 1973, Black and Scholes [1] applied the hedging argument developed by Merton [2] to derive the analytical expression of fixed-volatility option price. Financial mathematics applied to derivative pricing has made great progress from then on [3, 4]. Recently, more complex options have emerged in financial markets, which are collectively called exotic options [2, 5]. A kind of exotic option is the one attached some conditions to an ordinary option, taking barrier options for example: when the underlying price touches the barrier, the option contract will be activated, which is called knock-in option; when the underlying price touches the

---

✉ Chao Guo
chaog@lfnu.edu.cn

Qi Chen
20080142038@lfnu.edu.cn

Hong-tao Wang
wanghongtao@lfnu.edu.cn

[1] School of Economics and Management, Langfang Normal University, Langfang 065000, China

[2] School of Science, Langfang Normal University, Langfang 065000, China

barrier, the option contract is invalid, which is called knock-out option. Snyder had discussed down-and-out option in 1969 [6]. Baaquie et al discussed this kind of option by Hamiltonian method, and derived the corresponding analytical expressions [7]. Similar to one-dimensional infinite square potential well in quantum mechanics, the analytical expression for double-knock-out option price was also derived [8], which is in accordance with the result derived by mathematical calculations [9]. In addition, path integral and Hamiltonian approaches have been applied to the research of interest rate derivative pricing [10, 11]. Some works investigating step options appear in [12, 13, 15]. Exotic barrier options pricing could be studied in [14, 16]. Study on option pricing by path integral and Hamiltonian approaches can be found in [7, 8, 17, 18].

In this paper, we will discuss two kinds of barrier options, which are called proportional floating barrier step (PFBS) option and proportional floating double-barrier step (PFDBS) option, respectively. Floating barrier options are the barrier options with time dependent barrier levels which were discussed in [19, 20]. And the study on floating barrier option pricing by path-integral approach is in [21]. A step option could be also called a gradual knock-out (knock-in) option: when the underlying price touches and passes the barrier, the option contract is not invalid immediately, but the option knocks out (knocks in) gradually [12, 13]. A proportional step call is defined by its payoff [13]

$$e^{-V_0 t} \max(S_T - K, 0) \tag{1}$$

where $S_T$ is the underlying price on the expiration date, $K$ is the exercise price, $V_0$ is called knock-out rate, which is corresponding to the potential in quantum mechanics. The exponential $e^{-V_0 t}$ is interpreted as a knock-out discount factor with discounting time $t$. A daily knock-out factor $d$ could be defined as [12, 13]

$$d = e^{-V_0/250} \tag{2}$$

where 250 trading days per year is assumed. A proportional floating barrier step option is considered as a proportional step option combined with a floating barrier. We focus on the relation between proportional floating barrier step call (proportional floating double-barrier step call) option pricing and one dimensional trapezoid potential scattering (finite symmetric square well): when a particle moving ahead and going through the boundary, the wave function begins to decay exponentially, which is similar to a proportional floating barrier step (proportional floating double-barrier step) option knocks out over time.

Our work is organized as follows. In Sect. 2, we will derive the analytical expressions for proportional floating barrier step call option price by path integral approach. In Sect. 3, we will derive the analytical expressions for floating double-barrier option price by path integral approach. The analytical expressions for proportional floating double-barrier step call price are given in Sect. 4. In Sect. 5, we show the numerical results for option price as a function of underlying price, exercise price interest rate and floating rate, respectively. We summarize our main results in Sect. 6. The pricing formulas for Black–Scholes model, the standard barrier option (SBO), the standard double-barrier option (SDBO), proportional step option and proportional double-barrier step option derived by path integral method are reviewed in Appendix A–E.

## 2 Proportional Floating Barrier Step Option pricing

The partial differential equation (PDE) describes the price evolution for a proportional floating barrier step (PFBS) option is

$$\frac{\partial C}{\partial t} + rS\frac{\partial C}{\partial S} + \frac{1}{2}\sigma^2 S^2 \frac{\partial^2 C}{\partial S^2} - U(S,t)C = 0 \tag{3}$$

which is the same as the Black–Scholes equation except the barrier term in the left side. The barrier $U(S,t)$ is defined as [19]

$$U(S,t) = \begin{cases} r, & S \le S_0 e^{\delta t} \\ r + U_0, & S > S_0 e^{\delta t} \end{cases}, \quad t \in [0, \tau] \tag{4}$$

where $S(t)$ is the underlying asset price, $C(S(t),t)$ is the corresponding option price, $r$ is the fixed interest rate, $\sigma$ is the volatility of $S(t)$, $S_0$ is the barrier level at time $t = 0$, $\delta$ is the floating rate, which denotes the change rate of the barrier. Making the following variable substitution [7]

$$S = e^x, \ x \in (-\infty, +\infty) \tag{5}$$

the derivatives in (3) would make following transformation

$$\frac{\partial C}{\partial S} = e^{-x}\frac{\partial C}{\partial x}, \quad \frac{\partial^2 C}{\partial S^2} = e^{-2x}\left(\frac{\partial^2 C}{\partial x^2} - \frac{\partial C}{\partial x}\right) \tag{6}$$

and the Black–Scholes equation (3) could be rewritten into

$$\frac{\partial C}{\partial t} = \left[ -\frac{\sigma^2}{2}\frac{\partial^2}{\partial x^2} + \left(\frac{1}{2}\sigma^2 - r\right)\frac{\partial}{\partial x} + V(x,t) \right]C \tag{7}$$

which could be viewed as a Schrödinger-type equation. Without loss of generality, we set $\delta > 0$ throughout the following discussion. An up-and-out proportional floating barrier step option price changing could be analogous to a one-dimensional particle moving under the following potential

$$V(x,t) = \begin{cases} r, & x \le B_0 + \delta t, \\ r + V_0, & x > B_0 + \delta t. \end{cases} \tag{8}$$

where

$$S = e^{B(t)}, \ S_0 = e^{B_0}, \ t \in [0, \tau] \tag{9}$$

for $x > B_0 + \delta t$, the wave function decays as the distance increases in the case of the particle energy $E < V_0$, which is similar to an option touches a barrier at $x = B_0 + \delta t$ and knocks out gradually. The Hamiltonian for a proportional floating barrier step (PFBS) option is

$$H_{\text{PFBS}} = -\frac{\sigma^2}{2}\frac{\partial^2}{\partial x^2} + \left(\frac{1}{2}\sigma^2 - r\right)\frac{\partial}{\partial x} + V(x,t) \tag{10}$$

the potential $V(x,t)$ is time dependent, so is the Hamiltonian $H_{\text{PFBS}}$. According to (8), $V(x,t)$ is binary: when $x < B_0 + \delta t$, $V(x,t) \equiv r$, and $H_{\text{PFBS}}$ degenerated to Black–Scholes Hamiltonian; when $x > B_0 + \delta t$, $V(x,t) \equiv r + V_0$, which is also a constant for a fixed $V_0$. The Hamiltonian (10) is non-Hermitian, for $x > B_0 + \delta t$, considering the following transformation

$$H_{\text{PFBS}} = e^{\alpha x}H_{\text{eff}}e^{-\alpha x} = e^{\alpha x}\left(-\frac{\sigma^2}{2}\frac{\partial^2}{\partial x^2} + \gamma + V_0\right)e^{-\alpha x} \tag{11}$$

where

$$\alpha = \frac{1}{\sigma^2}\left(\frac{\sigma^2}{2} - r\right), \quad \gamma = \frac{1}{2\sigma^2}\left(\frac{\sigma^2}{2} + r\right)^2 \tag{12}$$

and

$$H_{\text{eff}} = -\frac{\sigma^2}{2}\frac{\partial^2}{\partial x^2} + \gamma + V_0 \tag{13}$$

is Hermitian, which could be named the effective Hamiltonian. When $x$ is in different intervals ($x < B_0 + \delta t$ or $x > B_0 + \delta t$), the values of the potential are constants ($r$ or $r + V_0$), therefore the effective Hamiltonian $H_{\text{eff}}$ is independent of time $t$ and floating rate $\delta$. The stationary state Schrödinger equation for option price under $H_{\text{eff}}$ is

$$\begin{cases} \dfrac{\mathrm{d}^2\phi}{\mathrm{d}x^2} + p_1^2\phi = 0, & x \le B_0 + \delta t, \\ \dfrac{\mathrm{d}^2\phi}{\mathrm{d}x^2} - p_2^2\phi = 0, & x > B_0 + \delta t. \end{cases} \tag{14}$$

where $\phi$ is the stationary state wave function of the option, $E$ is corresponding to the option energy under the potential (8), and

$$p_1^2 = \frac{2(E - \gamma)}{\sigma^2}, \quad p_2^2 = \frac{2(\gamma + V_0 - E)}{\sigma^2} \tag{15}$$

here the condition $E < V_0$ has been considered, and the range of $p_1^2$ and $p_2^2$ is

$$p_1^2 < \frac{2(V_0 - \gamma)}{\sigma^2}, \quad p_2^2 = \frac{2V_0}{\sigma^2} - p_1^2 > \frac{2\gamma}{\sigma^2} \tag{16}$$

The general solution for (14) is

$$\phi(x, t) = \begin{cases} e^{ip_1[x - (B_0 + \delta t)]} + A_1 e^{-ip_1[x - (B_0 + \delta t)]}, & x \le B_0 + \delta t, \\ A_2 e^{-p_2[x - (B_0 + \delta t)]}, & x > B_0 + \delta t. \end{cases} \tag{17}$$

where

$$A_1 = \frac{p_1 - ip_2}{p_1 + ip_2}, \quad A_2 = \frac{2p_1}{p_1 + ip_2} \tag{18}$$

here the boundary condition at $x = B_0 + \delta t$ is used. Now we calculate the price of up-and-out proportional floating barrier step call. Let $\tau_1$ indicates the occupation time below the barrier, and $\tau_2$ is the occupation time above the barrier. The pricing kernel is

$$\begin{aligned} p_{\text{PFBS}}(x, x'; \tau) &= \langle x|e^{-\tau_1 H_1 - \tau_2 H_2}|x'\rangle \\ &= e^{\alpha(x - x')} \int \frac{\mathrm{d}p}{2\pi} \langle x|e^{-\tau_1 H_{\text{eff1}}}|p\rangle\langle p|e^{-\tau_2 H_{\text{eff2}}}|x'\rangle \\ &= e^{\alpha(x - x')} \int \frac{\mathrm{d}p}{2\pi} e^{-\tau_1(\frac{1}{2}\sigma^2 p_1^2 + \gamma)}\langle x|p\rangle e^{-\tau_2(\gamma + V_0 - \frac{1}{2}\sigma^2 p_2^2)}\langle p|x'\rangle \\ &= e^{\alpha(x - x')} e^{-\tau\gamma} \int \frac{\mathrm{d}p}{2\pi} e^{-\frac{1}{2}\tau\sigma^2 p_1^2}\phi(x, 0)\phi^*(x', \tau) \end{aligned} \tag{19}$$

where $\phi^*(x, t)$ is the complex conjugate of $\phi(x, t)$, and the option price is

$$C_{\text{PFBS}}(x; \tau) = \int_{\ln K}^{+\infty} \mathrm{d}x' \, p_{\text{PFBS}}(x, x'; \tau)(e^{x'} - K) \tag{20}$$

where $K$ is the exercise price, $S = e^x$ is the initial underlying asset price, $S' = e^{x'}$ is the underlying asset price at $t = \tau$, and

$$
H_1 = -\frac{\sigma^2}{2}\frac{\partial^2}{\partial x^2} + \left(\frac{1}{2}\sigma^2 - r\right)\frac{\partial}{\partial x} + r, \quad H_2 = -\frac{\sigma^2}{2}\frac{\partial^2}{\partial x^2} + \left(\frac{1}{2}\sigma^2 - r\right)\frac{\partial}{\partial x} + r + V_0
$$

$$
H_{\text{eff1}} = -\frac{\sigma^2}{2}\frac{\partial^2}{\partial x^2} + \gamma, \quad H_{\text{eff2}} = -\frac{\sigma^2}{2}\frac{\partial^2}{\partial x^2} + \gamma + V_0
$$
(21)

We set the exercise price $K < e^{B_0}$. For $x < B_0 + \delta t$ and $x > B_0 + \delta t$, the wave functions are different, and the integral for (20) should be split into four cases

- $x < B_0$, and $\ln K < x' < B_0 + \delta\tau$ the pricing kernel

$$
p_{\text{PFBS1}}(x, x'; \tau)
$$

$$
= e^{-\tau\gamma} e^{\alpha(x-x')} \int_0^{\frac{\sqrt{2(V_0-\gamma)}}{\sigma}} \frac{\mathrm{d}p_1}{2\pi} e^{-\frac{1}{2}\tau\sigma^2 p_1^2} \left( e^{ip_1(x-B_0)} + A_1 e^{-ip_1(x-B_0)} \right)
$$

$$
\times \left( e^{-ip_1[x'-(B_0+\delta\tau)]} + A_1^* e^{ip_1[x'-(B_0+\delta\tau)]} \right)
$$

$$
= e^{-\tau\gamma} e^{\alpha(x-x')} \int_0^{\frac{\sqrt{2(V_0-\gamma)}}{\sigma}} \frac{\mathrm{d}p_1}{2\pi} e^{-\frac{1}{2}\tau\sigma^2 p_1^2} \left[ 2\cos p_1(x - x' + \delta\tau) + \frac{\sigma^2}{V_0}(p_1^2 - p_2^2) \right.
$$

$$
\left. \cos p_1(x + x' - 2B_0 - \delta\tau) - \frac{2\sigma^2}{V_0} p_1 p_2 \sin p_1(x + x' - 2B_0 - \delta\tau) \right]
$$
(22)

- $x < B_0$, and $x' > B_0 + \delta\tau$
  the pricing kernel

$$
p_{\text{PFBS2}}(x, x'; \tau)
$$

$$
= e^{-\tau\gamma} e^{\alpha(x-x')} \int_0^{\frac{\sqrt{2(V_0-\gamma)}}{\sigma}} \frac{\mathrm{d}p_1}{2\pi} e^{-\frac{1}{2}\tau\sigma^2 p_1^2}
$$

$$
\left( e^{ip_1(x-B_0)} + A_1 e^{-ip_1(x-B_0)} \right) A_2^* e^{-p_2[x'-(B_0+\delta\tau)]}
$$

$$
= 2e^{-\tau\gamma} e^{\alpha(x-x')} \int_0^{\frac{\sqrt{2(V_0-\gamma)}}{\sigma}} \frac{\mathrm{d}p_1}{2\pi} e^{-\frac{1}{2}\tau\sigma^2 p_1^2} \left[ \frac{\sigma^2}{V_0} p_1^2 \cos p_1(x - B_0 - \delta\tau) \right.
$$

$$
\left. - \frac{\sigma^2}{V_0} p_1 p_2 \sin p_1(x - B_0 - \delta\tau) \right] e^{-p_2(x'-B_0-\delta_\tau)}
$$
(23)

the option price for $x < B_0$ is

$$
C(x; \tau)|_{x<B_0} = \int_{\ln K}^{B_0+\delta\tau} p_{\text{PFBS1}}(x, x'; \tau)(e^{x'} - K) + \int_{B_0+\delta\tau}^{+\infty} p_{\text{PFBS2}}(x, x'; \tau)(e^{x'} - K)
$$
(24)

- $x > B_0$, and $\ln K < x' < B_0 + \delta\tau$
  the pricing kernel

$$p_{\text{PFBS3}}(x, x'; \tau)$$

$$= e^{-\tau\gamma} e^{\alpha(x-x')} \int_0^{\frac{\sqrt{2(V_0-\gamma)}}{\sigma}} \frac{\mathrm{d}p_1}{2\pi} e^{-\frac{1}{2}\tau\sigma^2 p_1^2} A_2 e^{-p_2(x-B_0)}$$

$$\left( e^{-ip_1[(x'-(B_0+\delta\tau)]} + A_1^* e^{ip_1[(x'-(B_0+\delta\tau)]} \right)$$

$$= 2e^{-\tau\gamma} e^{\alpha(x-x')} \int_0^{\frac{\sqrt{2(V_0-\gamma)}}{\sigma}} \frac{\mathrm{d}p_1}{2\pi} e^{-\frac{1}{2}\tau\sigma^2 p_1^2} e^{-p_2(x-B_0-\delta\tau)} \left[ \frac{\sigma^2}{V_0} p_1^2 \cos p_1(x'-B_0-\delta\tau) \right.$$

$$\left. - \frac{\sigma^2}{V_0} p_1 p_2 \sin p_1(x'-B_0-\delta\tau) \right]$$

$$\tag{25}$$

- $x > B_0$, and $x' > B_0 + \delta\tau$

$$p_{\text{PFBS4}}(x, x'; \tau)$$

$$= e^{-\tau\gamma} e^{\alpha(x-x')} \int_0^{\frac{\sqrt{2(V_0-\gamma)}}{\sigma}} \frac{\mathrm{d}p_1}{2\pi} e^{-\frac{1}{2}\tau\sigma^2 p_1^2} A_2 e^{-p_2(x-B_0)} A_2^* e^{-p_2[(x'-(B_0+\delta\tau)]} \tag{26}$$

$$= 2e^{-\tau\gamma} e^{\alpha(x-x')} \int_0^{\frac{\sqrt{2(V_0-\gamma)}}{\sigma}} \frac{\mathrm{d}p_1}{2\pi} e^{-\frac{1}{2}\tau\sigma^2 p_1^2} \frac{\sigma^2}{V_0} p_1^2 e^{-p_2(x+x'-2B_0-\delta\tau)}$$

the option price for $x > B_0$ is

$$C(x; \tau)|_{x>B_0} = \int_{\ln K}^{B_0+\delta\tau} p_{\text{PSO3}}(x, x'; \tau)(e^{x'} - K) + \int_{B_0+\delta\tau}^{+\infty} p_{\text{PSO4}}(x, x'; \tau)(e^{x'} - K) \tag{27}$$

## 3 Floating Double-Barrier Option Pricing

Consider the following barriers [19]

$$A = A_0, \quad B(t) = B_0 e^{\delta t}, \quad t \in [0, \tau] \tag{28}$$

where $A$ is the fixed lower bound ($A_0 < B_0$), $B(t)$ is the upper bound changing with time. We set $\delta > 0$, which means the width of barriers increases with time. Let

$$A = e^a, \quad B(t) = e^{b(t)} \tag{29}$$

and the barriers could be converted into

$$a = a_0, \quad b(t) = b_0 + \delta t, \quad t \in [0, \tau] \tag{30}$$

where $B_0 = e^{b_0}$. The potential for floating double barrier (FDB) option is

$$V(x) = \begin{cases} \infty, & x \leq a_0, \\ 0, & a_0 < x < b_0 + \delta t, \\ \infty, & x \geq b_0 + \delta t. \end{cases} \tag{31}$$

and the orthonormal eigenstate is

$$\phi_n(x) = \begin{cases} \sqrt{\dfrac{2}{b_0 - a_0 + \delta t}}\,\sin[p_n(x - a_0)], & a_0 < x < b_0 + \delta t, \\[3mm] 0, & x < a_0,\ x > b_0 + \delta t. \end{cases} \tag{32}$$

where

$$p_n = \frac{n\pi}{b_0 - a_0 + \delta t} \tag{33}$$

Owing to $p_n$ changes with time $t$, we discretize $\tau$ so that there are $N$ steps to maturity, with each time step is

$$\epsilon = \tau / N \tag{34}$$

The pricing kernel could be rewritten as

$$\begin{aligned} P_{\text{FDB}}(x, x'; \epsilon, \tau) &= \langle x | e^{-\tau H_{\text{FDB}}} | x' \rangle \\ &= \lim_{\epsilon \to \infty} \int_{a_0}^{b_0 + \delta\epsilon} dx_1 \int_{a_0}^{b_0 + 2\delta\epsilon} dx_2 ... \int_{a_0}^{b_0 + (N-1)\delta\epsilon} dx_{N-1} \\ &\quad \times \langle x | e^{-\epsilon H_{\text{FDB}}} | x_1 \rangle \langle x_1 | e^{-\epsilon H_{\text{FDB}}} | x_2 \rangle \cdots \langle x_{N-1} | e^{-\epsilon H_{\text{FDB}}} | x' \rangle \end{aligned} \tag{35}$$

where $H_{\text{DFB}}$ is the same as (11) except that $V(x)$ is defined by (31). Now we calculate $jth$ matrix element

$$\langle x_{j-1} | e^{-\epsilon H_{\text{DFB}}} | x_j \rangle = e^{\alpha(x_{j-1} - x_j)} e^{-\epsilon\gamma} \sum_{n=1}^{+\infty} e^{-\frac{1}{2}\epsilon\sigma^2 p_{n,j-1}^2} \phi_n(x_{j-1}) \phi_n(x_j) \tag{36}$$

where

$$\phi_n(x_j) = \sqrt{\frac{2}{b_0 - a_0 + j\delta\epsilon}}\,\sin\left[\frac{n\pi}{b_0 - a_0 + j\delta\epsilon}(x - a_0)\right] \tag{37}$$

similarly

$$\begin{aligned} &\langle x_{j-1} | e^{-2\epsilon H_{\text{DFB}}} | x_{j+1} \rangle \\ &= \int_{a_0}^{b_0 + jd\epsilon} dx_j \langle x_{j-1} | e^{-\epsilon H_{\text{DFB}}} | x_j \rangle \langle x_j | e^{-\epsilon H_{\text{DFB}}} | x_{j+1} \rangle \\ &= \int_{a_0}^{b_0 + jd\epsilon} dx_j\, e^{\alpha(x_{j-1} - x_{j+1})} e^{-2\epsilon\gamma} \sum_{n=1}^{+\infty} \sum_{n'=1}^{+\infty} e^{-\frac{1}{2}\epsilon\sigma^2 (p_{n,j-1}^2 + p_{n',j}^2)} \\ &\quad \times \phi_n(x_{j-1}) \phi_n(x_j) \phi_{n'}(x_j) \phi_{n'}(x_{j+1}) \\ &= e^{\alpha(x_{j-1} - x_{j+1})} e^{-2\epsilon\gamma} \sum_{n=1}^{+\infty} e^{-\frac{1}{2}\epsilon\sigma^2 n^2 \pi^2 \left(\frac{1}{(b_0 - a_0 + (j-1)\delta\epsilon)^2} + \frac{1}{(b_0 - a_0 + j\delta\epsilon)^2}\right)} \phi_n(x_{j-1}) \phi_n(x_{j+1}) \end{aligned} \tag{38}$$

where the orthonormalization condition

$$\int_{a_0}^{b_0 + jd\epsilon} dx_j\, \phi_n(x_j) \phi_{n'}(x_j) = \delta_{nn'} = \begin{cases} 0, & n \neq n', \\ 1, & n = n'. \end{cases} \tag{39}$$

has been used. After some similar calculation, the pricing kernel(35) could be denoted as

$$
\begin{aligned}
P_{\mathrm{DFB}}&(x, x'; \epsilon, \tau) \\
&= \langle x | e^{-\tau H_{\mathrm{FDB}}} | x' \rangle \\
&= e^{\alpha(x-x')} e^{-\tau \gamma} \sum_{n=1}^{+\infty} e^{-\frac{1}{2}\sigma^2 n^2 \pi^2 \lim_{N \to \infty} \epsilon \sum_{j=0}^{N-1} \frac{1}{(b_0 - a_0 + j\delta\epsilon)^2}} \\
&\quad \times \sqrt{\frac{2}{b_0 - a_0}} \sqrt{\frac{2}{b_0 - a_0 + \delta\tau}} \sin\left[\frac{n\pi}{b_0 - a_0}(x - a_0)\right] \sin\left[\frac{n\pi}{b_0 - a_0 + \delta\tau}(x' - a_0)\right]
\end{aligned}
\tag{40}
$$

where the summation in the exponential could be calculated as follow

$$
\begin{aligned}
\sum_{j=0}^{N-1} \epsilon \frac{1}{(b_0 - a_0 + j\delta\epsilon)^2} &= \frac{1}{\delta^2 \epsilon} \sum_{j=0}^{N-1} \frac{1}{\left(j + \frac{b_0 - a_0}{\delta\epsilon}\right)^2} \\
&= \frac{1}{\delta^2 \epsilon} \left[\sum_{j=0}^{+\infty} \frac{1}{\left(j + \frac{b_0 - a_0}{\delta\epsilon}\right)^2} - \sum_{j=N}^{+\infty} \frac{1}{\left(j + \frac{b_0 - a_0}{\delta\epsilon}\right)^2}\right] \\
&= \frac{1}{\delta^2 \epsilon} \left[\sum_{j=0}^{+\infty} \frac{1}{\left(j + \frac{b_0 - a_0}{\delta\epsilon}\right)^2} - \sum_{k=0}^{+\infty} \frac{1}{\left(k + \frac{b_0 - a_0}{\delta\epsilon} + N\right)^2}\right] \\
&= \frac{1}{\delta^2 \epsilon} \left[\psi'\left(\frac{b_0 - a_0}{\delta\epsilon}\right) - \psi'\left(\frac{b_0 - a_0}{\delta\epsilon} + N\right)\right]
\end{aligned}
\tag{41}
$$

where

$$
\psi(z) = \frac{\Gamma'(z)}{\Gamma(z)}
\tag{42}
$$

is the $\psi$ function, and

$$
\Gamma(z) = \int_0^\infty e^{-t} t^{z-1} \mathrm{d}t, \quad \mathrm{Re}\, z > 0
\tag{43}
$$

is the $\Gamma$ function.

## 4 Proportional Floating Double-Barrier Step Option Pricing

The price changing of a proportional floating double-barrier step (PFDBS) option could be analogous to a particle moving in a symmetric square potential well with the potential

$$
V(x) = \begin{cases} 0, & a_0 < x < b_0 + \delta t, \\ V_0, & x \le a_0,\ x \ge b_0 + \delta t. \end{cases}
\tag{44}
$$

where the lower barrier is fixed at $a_0$, and the upper barrier level increases with time ($\delta > 0$). For $x \le a_0$ or $x \ge b_0 + \delta t$, the wave function decays with the increasing distances from the well, which is similar to an option touches a barrier and knocks out gradually. The Hamiltonian $H_{\mathrm{PFDBS}}$ is the same as (11) except that $V(x)$ is defined by (44). The stationary state Schrödinger equation for option price is

$$\begin{cases} -\dfrac{\sigma^2}{2}\dfrac{\mathrm{d}^2\phi}{\mathrm{d}x^2} + \gamma\phi = E\phi, & a_0 < x < b_0 + \delta t, \\ -\dfrac{\sigma^2}{2}\dfrac{\mathrm{d}^2\phi}{\mathrm{d}x^2} + (\gamma + V_0)\phi = E\phi, & x < a_0,\ x > b_0 + \delta t. \end{cases} \tag{45}$$

where $\phi$ is the option price wave function, $E$ is corresponding to bound state energy levels in the potential well. (45) could be simplified into

$$\begin{cases} \dfrac{\mathrm{d}^2\phi}{\mathrm{d}x^2} + k_1^2\phi = 0, & a_0 < x < b_0 + \delta t, \\ \dfrac{\mathrm{d}^2\phi}{\mathrm{d}x^2} - k_2^2\phi = 0, & x < a_0,\ x > b_0 + \delta t. \end{cases} \tag{46}$$

where

$$k_1^2 = \frac{2(E - \gamma)}{\sigma^2}, \quad k_2^2 = \frac{2(V_0 + \gamma - E)}{\sigma^2} \tag{47}$$

the corresponding eigenwave function is

$$\phi(x,t) = \begin{cases} A_3\ e^{k_2\left(x - \frac{b_0 + a_0 + \delta t}{2}\right)}, & x \le a_0, \\ A_1 \sin(k_1 x + \xi), & a_0 t < x \le b_0 + \delta t, \\ A_2\ e^{-k_2\left(x - \frac{b_0 + a_0 + \delta t}{2}\right)}, & x > b_0 + \delta t. \end{cases} \tag{48}$$

Considering the continuity for both wave function and its derivative at $x = a_0$ and $x = b_0 + \delta t$, we have

$$\delta = \frac{\ell\pi}{2} - k_1\frac{b_0 + a_0 + \delta t}{2}, \quad \ell = 0, 1, 2, \dots \tag{49}$$

and (48) could be split into two parts

$$\phi_1(x,t) = \begin{cases} -A_{21}e^{k_2\left(x - \frac{b_0 + a_0 + \delta t}{2}\right)}, & x \le a_0, \\ A_1 \sin k_1\left(x - \dfrac{b_0 + a_0 + \delta t}{2}\right), & a_0 \le x \le b_0 + \delta t, \quad for\ \ell = 0, 2, 4, \dots \\ A_{21}e^{-k_2\left(x - \frac{b_0 + a_0 + \delta t}{2}\right)}, & x > b_0 + \delta t. \end{cases} \tag{50}$$

and

$$\phi_2(x,t) = \begin{cases} A_{22}e^{k_2\left(x - \frac{b_0 + a_0 + \delta t}{2}\right)}, & x \le a_0, \\ A_1 \cos k_1\left(x - \dfrac{b_0 + a_0 + \delta t}{2}\right), & a_0 \le x \le b_0 + \delta t, \quad for\ \ell = 1, 3, 5, \dots \\ A_{22}e^{-k_2\left(x - \frac{b_0 + a_0 + \delta t}{2}\right)}, & x > b_0 + \delta t. \end{cases} \tag{51}$$

where

$$A_1 = \sqrt{\frac{2k_2}{k_2(b_0 - a_0 + \delta t) + 2}}, \quad A_{21} = A_1 \sin\frac{k_1(b_0 - a_0 + \delta t)}{2}e^{k_2\frac{b_0 - a_0 + \delta t}{2}},$$
$$A_{22} = A_1 \cos\frac{k_1(b_0 - a_0 + \delta t)}{2}e^{k_2\frac{b_0 - a_0 + \delta t}{2}} \tag{52}$$

the normalization condition has been used. Considering boundary conditions for (50) and (51) at $x = b_0 + \delta t$ respectively, we have

$$\cot \frac{k_1(b_0 - a_0 + \delta t)}{2} = -\frac{k_2}{k_1} \tag{53}$$

$$\tan \frac{k_1(b_0 - a_0 + \delta t)}{2} = \frac{k_2}{k_1} \tag{54}$$

let

$$\beta = \sqrt{k_1^2 + k_2^2} = \frac{\sqrt{2V_0}}{\sigma} \tag{55}$$

(53) and (54) could be combined into

$$k_{1n} \frac{b_0 - a_0 + \delta t}{2} = \frac{n\pi}{2} - \arcsin \frac{k_{1n}}{\beta}, \quad n = 1, 2, 3, \dots \tag{56}$$

(56) is the energy level equation. In general, there is no analytical solution for energy eigenvalues, yet the approximate expression could be derived. For low energy levels ($k_{1n} \ll \beta$), $\arcsin(k_{1n}/\beta) \approx k_{1n}/\beta$, (56) could be simplified into

$$k_{1n,\text{low}} \approx \frac{\beta n\pi}{\beta(b_0 - a_0 + \delta t) + 2} \tag{57}$$

for high energy levels ($k_{1n} \approx \beta$), $\arcsin(1 - k_{1n}/\beta) \approx \pi/2 - \sqrt{2k_{1n}/\beta}$, (56) could be simplified into

$$k_{1n,\text{high}} \frac{b_0 - a_0 + \delta t}{2} \approx \frac{(n-1)\pi}{2} + \sqrt{2 - \frac{2(n-1)\pi}{\beta(b_0 - a_0 + \delta t)}} \tag{58}$$

Now we calculate the expressions of option pricing kernels. The maximum number of energy levels is determined by

$$n_{\max} = \left[ \frac{\beta(b_0 - a_0 + \delta t)}{\pi} \right] \tag{59}$$

the bracket [ ] represents the minimal integer not less than $\beta(b_0 - a_0 + \delta t)/\pi$. (59)indicates that $n_{\max}$ increases with time. We assume that there are $n_0$ energy levels during $(0, \tau_1)$, $(n_0 + 1)$ energy levels during $(\tau_1, \tau_2)$,..., $(n_0 + N)$ energy levels during $(\tau_N, \tau)$, and

$$0 < \tau_1 < \tau_2 < \cdots < \tau_N < \tau \tag{60}$$

the pricing kernel could be denoted as

$$\langle x | e^{-\tau H_{\text{PFDBS}}} | x' \rangle$$
$$= \int_{-\infty}^{+\infty} dx_1 \int_{-\infty}^{+\infty} dx_2 \cdots \int_{-\infty}^{+\infty} dx_N \langle x | e^{-\tau_1 H_{\text{PFDBS}}} | x_1 \rangle \langle x_1 | e^{-(\tau_2 - \tau_1) H_{\text{PFDBS}}} | x_2 \rangle \times \quad (61)$$
$$\cdots \times \langle x_N | e^{-(\tau - \tau_N) H_{\text{PFDBS}}} | x' \rangle$$

according to (40), the pricing kernel for $(\tau_i, \tau_{i+1})$ is

$$
\begin{aligned}
&\langle x_i | e^{-(\tau_{i+1}-\tau_i)H_{\text{PFDBS}}} | x_{i+1} \rangle \\
&= e^{\alpha(x_{i+1}-x_i)} e^{-(\tau_{i+1}-\tau_i)\gamma} \sum_{n=1}^{n_0+i} e^{-\frac{1}{2}\sigma^2 \lim\limits_{\epsilon \to 0} \epsilon \sum\limits_{j=0}^{(\tau_{i+1}-\tau_i)/\epsilon} (k_{1n,j\epsilon})^2} \phi(x_1, \tau_i)\phi(x_2, \tau_{i+1}) \\
&= e^{\alpha(x_{i+1}-x_i)} e^{-(\tau_{i+1}-\tau_i)\gamma} \left[ \sum_{n=2,4,\ldots} e^{-\frac{1}{2}\sigma^2 \lim\limits_{\epsilon \to 0} \epsilon \sum\limits_{j=0}^{(\tau_{i+1}-\tau_i)/\epsilon} (k_{1n,j\epsilon})^2} \phi_1(x_1, \tau_i)\phi_1(x_2, \tau_{i+1}) \right.\\
&\quad \left. + \sum_{n=1,3,\ldots} e^{-\frac{1}{2}\sigma^2 \lim\limits_{\epsilon \to 0} \epsilon \sum\limits_{j=0}^{(\tau_{i+1}-\tau_i)/\epsilon} (k_{1n,j\epsilon})^2} \phi_2(x_1, \tau_i)\phi_2(x_2, \tau_{i+1}) \right]
\end{aligned}
\tag{62}
$$

where $n_0$ is the number of energy levels for $(0, \tau_1)$, $\phi_1(x, t)$ and $\phi_2(x, t)$ are given by (50) and (51), respectively. The definition of tiny time interval $\epsilon$ is the same as (34). For low energy levels, the summation in the exponent during an arbitrary time interval $(\tau_i, \tau_{i+1})$ could be calculated as

$$
\begin{aligned}
\epsilon \sum_{j=0}^{(\tau_{i+1}-\tau_i)/\epsilon} (k_{1n,j\epsilon,\text{low}})^2 &= \beta^2 n^2 \pi^2 \epsilon \sum_{j=0}^{(\tau_{i+1}-\tau_i)/\epsilon} \frac{1}{[\beta(b_0 - a_0 + \delta\tau_i + j\delta\epsilon) + 2]^2} \\
&= \frac{n^2\pi^2}{\delta^2\epsilon} \sum_{j=0}^{(\tau_{i+1}-\tau_i)/\epsilon} \frac{1}{\left[ j + \frac{\beta(b_0 - a_0 + \delta\tau_i)+2}{\beta\delta\epsilon} \right]^2} \\
&= \frac{n^2\pi^2}{\delta^2\epsilon} \left[ \psi'\left( \frac{\beta(b_0 - a_0 + \delta\tau_i) + 2}{\beta\delta\epsilon} \right) \right. \\
&\quad \left. - \psi'\left( \frac{\beta(b_0 - a_0 + \delta\tau_i) + 2}{\beta\delta\epsilon} + \frac{\tau_{i+1} - \tau_i}{\epsilon} + 1 \right) \right]
\end{aligned}
\tag{63}
$$

for high energy levels, the expression is

$$
\begin{aligned}
\epsilon \sum_{j=0}^{N-1} (k_{1n,j\epsilon,\text{high}})^2 &= \epsilon \sum_{j=0}^{(\tau_{i+1}-\tau_i)/\epsilon} \left[ \frac{2}{b_0 - a_0 + \delta\tau_i + j\delta\epsilon} \left( \frac{(n-1)\pi}{2} \right.\right.\\
&\quad \left.\left. + \sqrt{2 - \frac{2(n-1)\pi}{\beta(b_0 - a_0 + \delta\tau_i + j\delta\epsilon)}} \right) \right]^2
\end{aligned}
\tag{64}
$$

in fact, the contribution to option price from large $n$ is negligible. We will only take low energy levels into account in the following section. The option price could be denoted as

$$
C_{\text{PFDBS}}(x; \tau_1, \tau_2, \ldots, \tau) = \int_{\ln K}^{+\infty} dx' \langle x | e^{-\tau H_{\text{PFDBS}}} | x' \rangle (e^{x'} - K)
\tag{65}
$$

where $K$ is the exercise price.

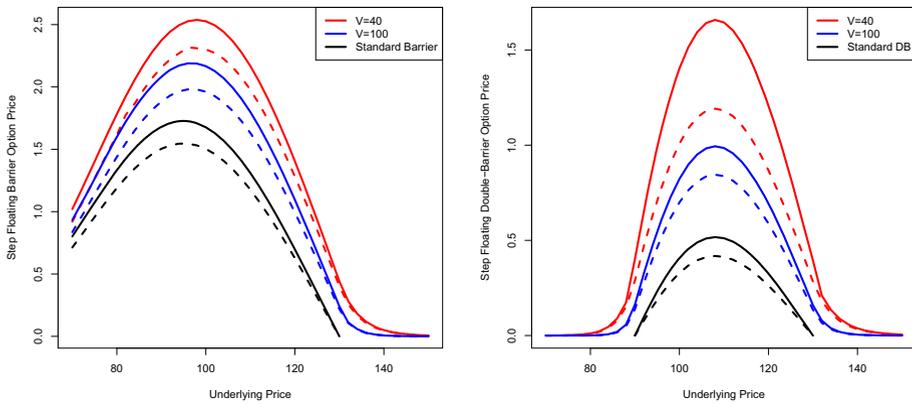

**Fig. 1** Proportional floating barrier step call price (left) and proportional floating double-barrier step call price (right) as functions of underlying asset price for different potentials, respectively. The dashed lines indicate the corresponding fixed barrier cases for comparison. Parameters: $a = \ln 100 = 4.5$, $b = \ln 130 = 4.867$, $S = 100$, $r = 0.05$, $\sigma = 0.3$, $\tau = 1$, $\delta = 0.01$

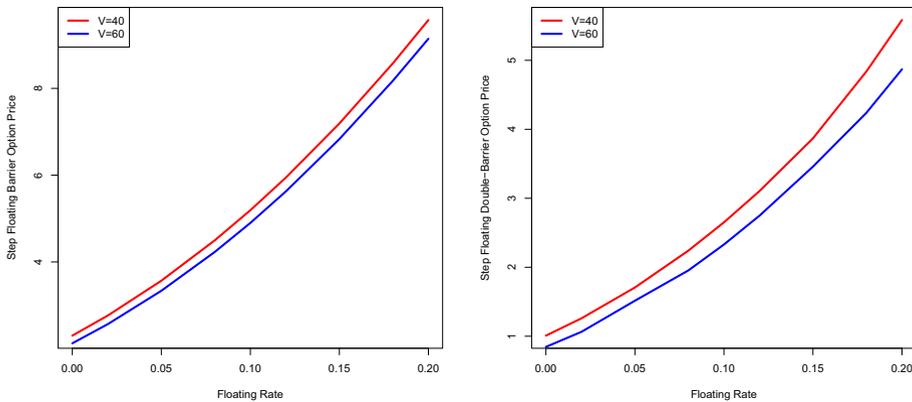

**Fig. 2** Proportional floating barrier step call price (left) and proportional floating double-barrier step call price (right) as functions of floating rate $\delta$ for different potentials, respectively. Parameters: $a = \ln 100 = 4.5$, $b = \ln 130 = 4.867$, $S = 100$, $r = 0.05$, $\sigma = 0.3$, $K = 100$, $\tau = 1$

## 5 Numerical Results

In Fig. 1, we show the proportional floating barrier step call price and the proportional floating double-barrier step call price as functions of underlying price, respectively, where the floating rate $\delta$ is set to 0.01. The dashed lines are corresponding to fixed barrier cases for comparison. It is shown that the option prices decrease with the increasing of potential $V_0$ for both the two diagrams. In the limit $V_0 \to \infty$, the option payoff tends to be the payoff of a standard step option (left) or of a standard double-barrier step option.

In Fig. 2, we show the proportional floating barrier step call price (left) and the proportional floating double-barrier step call price (right) as functions of floating rate $\delta$. For a fixed $V_0$, the option price increases with the increasing of the floating rate $\delta$.

In Fig. 3, we show the proportional floating barrier step call price (left) and the proportional floating double-barrier step call price (right) as functions of exercise price. The dashed lines

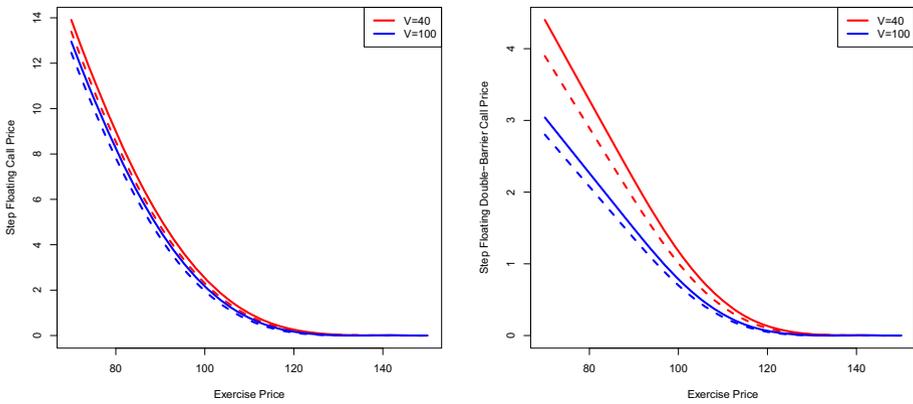

**Fig. 3** Proportional floating barrier step call price (left) and proportional floating double-barrier step call price (right) as functions of exercise price. Parameters: $a = \ln 100 = 4.5$, $b = \ln 130 = 4.867$, $S = 100$, $r = 0.05$, $\sigma = 0.3$, $\tau = 1$, $\delta = 0.01$

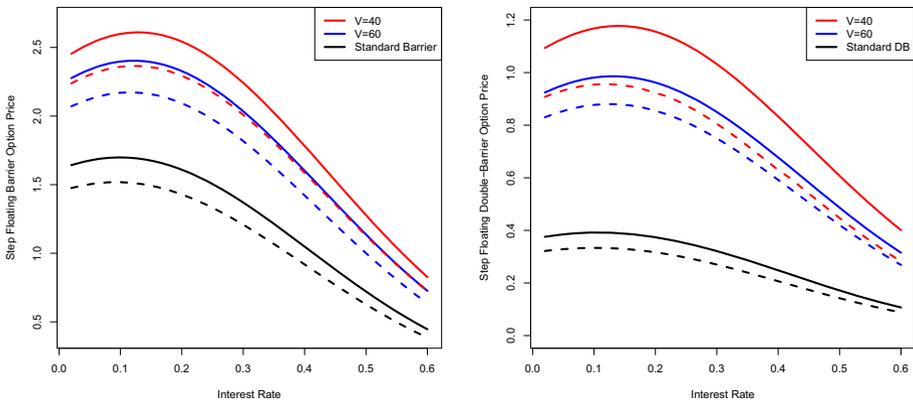

**Fig. 4** Proportional floating barrier step call price (left) and proportional floating double-barrier step call price (right) as functions of fixed interest rate. Parameters: $a = \ln 100 = 4.5$, $b = \ln 130 = 4.867$, $S = 100$, $\sigma = 0.3$, $\tau = 1$, $K = 100$, $\delta = 0.01$

are corresponding to fixed barrier cases for comparison. It is shown that the option prices decrease with the increasing of exercise price.

In Fig. 4, we show the proportional floating barrier step call price (left) and the proportional floating double-barrier step call price (right) as functions of fixed interest rate. The dashed lines are corresponding to fixed barrier cases for comparison. In the limit $V_0 \to \infty$, the option payoff tends to be the payoff of a standard step option (left) or of a standard double-barrier step option.

In Fig. 5, we show the proportional floating barrier step call delta (left) and the proportional floating double barrier step call delta (right) as functions of the underlying price. The definition of delta is

$$\Delta = \frac{\partial C}{\partial S} = e^{-x} \frac{\partial C}{\partial x} \tag{66}$$

the dashed lines are corresponding to the fixed barrier cases for comparison, respectively.

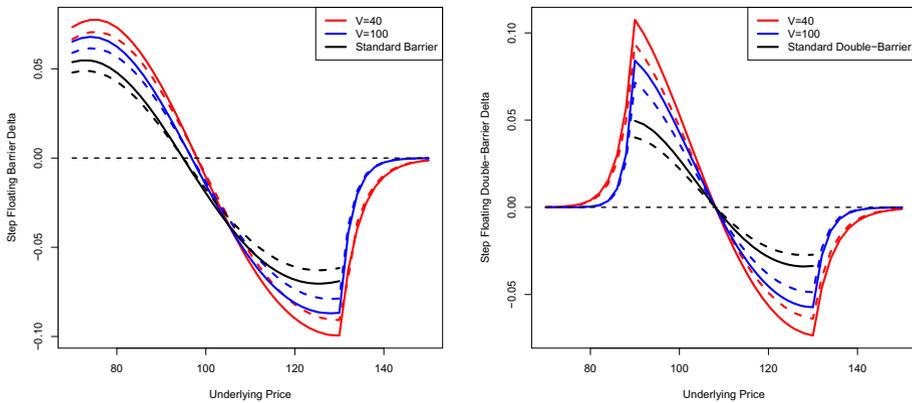

**Fig. 5** Proportional floating barrier step call price (left) and proportional floating double-barrier step call price (right) as functions of underlying price. Parameters: $a = \ln 100 = 4.5$, $B = b = \ln 130 = 4.867$, $S = 100$, $K = 100$, $r = 0.05$, $\sigma = 0.3$, $\tau = 1$, $\delta = 0.01$

## 6 Conclusion

The Hamiltonian approach is effective in linking option price changing to a particle moving under some potential in the space. Here we have studied pricing of the proportional floating barrier step call option and the proportional floating double-barrier step call option, which could be analogous to a particle moving through a trapezoid potential barrier or a symmetric square potential well, respectively. We have presented option prices changing with the initial underlying prices, floating rates, interest rates and exercise prices, respectively. The numerical results are in accordance with the results using mathematical approach in [12, 13]. The pricing of other barrier options could be studied by defining appropriate potentials $V$.

**Acknowledgements** Chao Guo and Hongtao Wang are supported by the Natural Science Foundation of Hebei Province under Grant No. A2022408002.

**Data Availability** The authors confirm that the data supporting the findings of this study are available within the article.

## Declaration

**Competing interests** The authors declare that they have no known competing financial interests or personal relationships that could have appeared to influence the work reported in this paper.

## Appendix A: The Black–Scholes Equation

In this section, we firstly illustrate how to derive the partial differential equation (PDE) for the price of a European call option using the riskless hedging principle [1]. Consider the following portfolio

$$\Pi = C - \Delta S \tag{A1}$$

which could be written into a differential form

$$d\Pi = dC - \Delta dS \tag{A2}$$

$\Pi$ is a portfolio in which an investor holds an option $C$ and short sells $\Delta$ amount of underlying asset $S$. The underlying asset price $S$ is assumed to follow the Geometric Brownian motion

$$\mathrm{d}S = \mu S \mathrm{d}t + \sigma S \mathrm{d}z \tag{A3}$$

where $\mu$ is the expected return rate, $\sigma$ is the asset volatility, and $z$ is the standard Brownian process. In the frame of Black–Scholes model, $\mu$ and $\sigma$ are assumed to be constant. $C$ is the function of $S$ and time $t$, using Ito lemma [22], we have

$$\mathrm{d}C = \frac{\partial C}{\partial t}\mathrm{d}t + \frac{1}{2}\sigma^2 S^2 \frac{\partial^2 C}{\partial S^2}\mathrm{d}t + \frac{\partial C}{\partial S}\mathrm{d}S \tag{A4}$$

By virtue of no arbitrage, the rate of return on $\Pi$ must equal the risk-free return given by the risk-free interest rate $r$, that is

$$\frac{\mathrm{d}\Pi}{\Pi} = r\mathrm{d}t \tag{A5}$$

substituting (A3) (A4) and (A5) into (A2), we obtain

$$\left(\frac{\partial C}{\partial t} + \frac{1}{2}\sigma^2 S^2 \frac{\partial^2 C}{\partial S^2} + \mu S \frac{\partial C}{\partial S} - \Delta \mu S\right)\mathrm{d}t + \left(\sigma S \frac{\partial C}{\partial S} - \Delta \sigma S\right)\mathrm{d}z = r(C - \Delta S)\mathrm{d}t \tag{A6}$$

since the right side of (A6) is risk-free, the coefficient of the stochastic term must be zero, and we have the amount of $S$

$$\Delta = \frac{\partial C}{\partial S} \tag{A7}$$

substituting (A7) into (A6), the Black–Scholes PDE could be denoted as

$$\frac{\partial C}{\partial t} + \frac{1}{2}\sigma^2 S^2 \frac{\partial^2 C}{\partial S^2} + rS\frac{\partial C}{\partial S} - rC = 0 \tag{A8}$$

consider the substitution

$$S = e^x, \quad -\infty < x < +\infty \tag{A9}$$

the Black–Scholes PDE could be rewritten into a Schrödinger-type equation

$$\frac{\partial C}{\partial t} = \left[-\frac{\sigma^2}{2}\frac{\partial^2}{\partial x^2} + \left(\frac{1}{2}\sigma^2 - r\right)\frac{\partial}{\partial x} + r\right]C \tag{A10}$$

with the Hamiltonian

$$H = -\frac{\sigma^2}{2}\frac{\partial^2}{\partial x^2} + \left(\frac{1}{2}\sigma^2 - r\right)\frac{\partial}{\partial x} + r \tag{A11}$$

## Appendix B: Hamiltonian Approach to Black–Scholes Model Pricing

According to Ref [8], starting from Black–Scholes pricing formula, the price of European option can be derived by path integral method. The Black–Scholes formula is

$$\frac{\partial C}{\partial t} + rS\frac{\partial C}{\partial S} + \frac{1}{2}\sigma^2 S^2 \frac{\partial^2 C}{\partial S^2} - rC = 0 \tag{B1}$$

where $C$ is European option price, $S$ is the underlying asset price, $\sigma$ is the fixed volatility, and $r$ is the interest rate. Let

$$S = e^x, \quad (-\infty < x < +\infty) \tag{B2}$$

and (B1) can be denoted as

$$\frac{\partial C}{\partial t} = \left[ -\frac{\sigma^2}{2}\frac{\partial^2}{\partial x^2} + \left(\frac{1}{2}\sigma^2 - r\right)\frac{\partial}{\partial x} + r \right] C \tag{B3}$$

let

$$H_{\text{BS}} = -\frac{\sigma^2}{2}\frac{\partial^2}{\partial x^2} + \left(\frac{1}{2}\sigma^2 - r\right)\frac{\partial}{\partial x} + r \tag{B4}$$

the Black–Scholes equation is written as

$$\frac{\partial C}{\partial t} = H_{\text{BS}} C \tag{B5}$$

Comparing (B5) to Schrödinger equation, we have

$$\sigma^2 \sim \frac{1}{m^2}, \quad C \sim \psi(x) \tag{B6}$$

where $m$ is the particle mass, and $\psi(x)$ is the wave function. The Black–Scholes Hamiltonian (B4) in momentum representation can be denoted as

$$H_{\text{BS}} = \frac{1}{2}\sigma^2 p^2 + i\left(\frac{1}{2}\sigma^2 - r\right)p + r \tag{B7}$$

where $p = -i\frac{\partial}{\partial x}$. The pricing kernel is

$$
\begin{aligned}
\langle x|e^{-\tau H_{\text{BS}}}|x'\rangle &= \int_{-\infty}^{+\infty} \frac{\mathrm{d}p}{2\pi} \langle x|e^{-\tau H_{\text{BS}}}|p\rangle\langle p|x'\rangle \\
&= e^{-r\tau}\int_{-\infty}^{+\infty} \frac{\mathrm{d}p}{2\pi} e^{-\frac{1}{2}\tau\sigma^2\left(p - \frac{x'-x_0}{\tau\sigma^2}\right)^2 - \frac{(x'-x_0)^2}{2\tau\sigma^2}} \\
&= \frac{1}{\sqrt{2\pi\tau\sigma^2}} e^{-r\tau} e^{-\frac{1}{2\tau\sigma^2}(x'-x_0)^2}
\end{aligned} \tag{B8}
$$

where the completeness relation has been used, and

$$x_0 = x + \tau\left(r - \frac{\sigma^2}{2}\right) \tag{B9}$$

The European call option price can be denoted as

$$
\begin{aligned}
C(x, \tau) &= e^{-r\tau}\int_{-\infty}^{+\infty} \frac{\mathrm{d}x'}{\sqrt{2\pi\tau\sigma^2}}(e^{x'} - K)_+ e^{-\frac{1}{2\tau\sigma^2}(x'-x_0)^2} \\
&= e^{-r\tau}\int_{\ln K - x_0}^{+\infty} \frac{\mathrm{d}x'}{\sqrt{2\pi\tau\sigma^2}}(e^{x'+x_0} - K) e^{-\frac{1}{2\tau\sigma^2}x'^2} \\
&= SN(d_+) - e^{-r\tau}KN(d_-)
\end{aligned} \tag{B10}
$$

where

$$N(x) = \frac{1}{\sqrt{2\pi}}\int_{-\infty}^{x} e^{-\frac{1}{2}z^2}\mathrm{d}z, \quad d_\pm = \frac{\ln\frac{S}{K} + \left(r \pm \frac{\sigma^2}{2}\right)\tau}{\sigma\sqrt{\tau}} \tag{B11}$$

## Appendix C: Hamiltonian Approach to the Standard Barrier Option Pricing

The up-and-out standard barrier (UOSB) option Hamiltonian is

$$
\begin{aligned}
H_{\text{UOSB}} &= H_{\text{BS}} + V(x) \\
&= -\frac{\sigma^2}{2}\frac{\partial^2}{\partial x^2} + \left(\frac{\sigma^2}{2} - r\right)\frac{\partial}{\partial x} + r + V(x) \\
&= e^{\alpha x}\left(-\frac{\sigma^2}{2}\frac{\partial^2}{\partial x^2} + \gamma\right)e^{-\alpha x} + V(x)
\end{aligned}
\tag{C1}
$$

where

$$
\alpha = \frac{1}{\sigma^2}\left(\frac{\sigma^2}{2} - r\right), \ \ \gamma = \frac{1}{2\sigma^2}\left(\frac{\sigma^2}{2} + r\right)^2
\tag{C2}
$$

and the potential $V(x)$ is

$$
V(x) = \begin{cases} 0, & x < B, \\ \infty, & x \geq B. \end{cases}
\tag{C3}
$$

the corresponding wave function is

$$
C(x) = \begin{cases} e^{ip(x-B)} - e^{-ip(x-B)}, & x < B, \\ 0, & x \geq B. \end{cases}
\tag{C4}
$$

and the pricing kernel is

$$
\begin{aligned}
&p_{\text{UOSB}}(x, x'; \tau) \\
&= \langle x|e^{-\tau H_{DB}}|x'\rangle \\
&= e^{-\tau\gamma}e^{\alpha(x-x')}\int_0^\infty \frac{\mathrm{d}p}{2\pi}e^{-\frac{1}{2}\tau\sigma^2 p^2}\left[e^{ip(x-B)} - e^{-ip(x-B)}\right]\left[e^{-ip(x'-B)} - e^{ip(x'-B)}\right] \\
&= 2e^{-\tau\gamma}e^{\alpha(x-x')}\int_0^\infty \frac{\mathrm{d}p}{2\pi}e^{-\frac{1}{2}\tau\sigma^2 p^2}[\cos(p(x - x')) - \cos(p(x + x' - 2B))]
\end{aligned}
\tag{C5}
$$

the corresponding option price

$$
\begin{aligned}
C_{\text{UOSB}}(x; \tau) &= \int_{\ln K}^B \mathrm{d}x'\, p_{\text{UOSB}}(x, x'; \tau)(e^{x'} - K) \\
&= 2e^{-\tau\gamma}\int_{\ln K}^B \mathrm{d}x'\, e^{\alpha(x-x')}\int_0^\infty \frac{\mathrm{d}p}{2\pi}e^{-\frac{1}{2}\tau\sigma^2 p^2} \\
&\quad [\cos(p(x - x')) - \cos(p(x + x' - 2B))](e^{x'} - K)
\end{aligned}
\tag{C6}
$$

# Appendix D: Hamiltonian Approach to the Standard Double-Barrier Option Pricing

The standard double-barrier (SDB) option Hamiltonian is [8]

$$
\begin{aligned}
H_{\text{SDB}} &= H_{\text{BS}} + V(x) \\
&= -\frac{\sigma^2}{2}\frac{\partial^2}{\partial x^2} + \left(\frac{\sigma^2}{2} - r\right)\frac{\partial}{\partial x} + r + V(x) \\
&= e^{\alpha x}\left(-\frac{\sigma^2}{2}\frac{\partial^2}{\partial x^2} + \gamma\right)e^{-\alpha x} + V(x)
\end{aligned}
\tag{D1}
$$

where

$$
\alpha = \frac{1}{\sigma^2}\left(\frac{\sigma^2}{2} - r\right), \ \ \gamma = \frac{1}{2\sigma^2}\left(\frac{\sigma^2}{2} + r\right)^2
\tag{D2}
$$

and the potential $V(x)$ is

$$
V(x) = \begin{cases} \infty, & x \le a, \\ 0, & a < x < b, \\ \infty, & x \ge b. \end{cases}
\tag{D3}
$$

the corresponding eigenstate is

$$
\phi_n(x) = \begin{cases} \sqrt{\dfrac{n\pi}{b-a}}\sin[p_n(x-a)], & a < x < b, \\ 0, & x < a, \ x > b. \end{cases}
\tag{D4}
$$

where

$$
p_n = \frac{n\pi}{b-a}, \ \ E_n = \frac{1}{2}\sigma^2 p_n^2, \ \ n = 1, 2, 3, \dots
\tag{D5}
$$

The pricing kernel is

$$
\begin{aligned}
p_{\text{SDB}}(x, x'; \tau) &= \langle x|e^{-\tau H_{DB}}|x'\rangle \\
&= e^{\alpha(x-x')}\langle x|e^{-\tau\left(-\frac{\sigma^2}{2}\frac{\partial^2}{\partial x^2} + \gamma + V\right)}|x'\rangle \\
&= e^{-\tau\gamma}e^{\alpha(x-x')}\sum_{n=1}^{+\infty}e^{-\frac{1}{2}\tau\sigma^2 p_n^2}\phi_n(x)\phi_n(x')
\end{aligned}
\tag{D6}
$$

and the option price

$$
\begin{aligned}
C_{\text{SDB}}(x; \tau) &= \int_{\ln K}^{b} dx'\, p_{\text{SDB}}(x, x'; \tau)(e^{x'} - K) \\
&= \frac{2}{b-a}e^{-\tau\gamma}\int_{\ln K}^{b} dx'\, e^{\alpha(x-x')} \\
&\quad \times \sum_{n=1}^{+\infty}e^{-\frac{1}{2}\tau\sigma^2\frac{n^2\pi^2}{(b-a)^2}}\sin\frac{n\pi(x-a)}{b-a}\sin\frac{n\pi(x'-a)}{b-a}(e^{x'} - K)
\end{aligned}
\tag{D7}
$$

## Appendix E: Hamiltonian Approach to Proportional Step Option Pricing

An up-and-out proportional step option price changing could be analogous to a one-dimensional particle moving under the following potential

$$V(x) = \begin{cases} 0, & x < B, \\ V_0, & x \geq B. \end{cases} \tag{E1}$$

for $x \leq B$, the wave function decays as the distance increases in the case of the particle energy $E < V_0$, which is similar to an option touches a barrier at $x = B$ and knocks out gradually. Making the following variable substitution [7]

$$S = e^x, \ x \in (-\infty, +\infty) \tag{E2}$$

the Hamiltonian for a proportional step option (PSO) is

$$H_{\mathrm{PSO}} = -\frac{\sigma^2}{2}\frac{\partial^2}{\partial x^2} + \left(\frac{1}{2}\sigma^2 - r\right)\frac{\partial}{\partial x} + r + V(x) \tag{E3}$$

where $S$ is the underlying asset price, $r$ is the fixed interest rate, and $\sigma$ is the volatility. The Hamiltonian (E3) is non-Hermitian, considering the following transformation

$$H_{\mathrm{PSO}} = e^{\alpha x} H_{\mathrm{eff}} e^{-\alpha x} = e^{\alpha x}\left(-\frac{\sigma^2}{2}\frac{\partial^2}{\partial x^2} + \gamma\right)e^{-\alpha x} + V(x) \tag{E4}$$

where

$$\alpha = \frac{1}{\sigma^2}\left(\frac{\sigma^2}{2} - r\right), \ \ \gamma = \frac{1}{2\sigma^2}\left(\frac{\sigma^2}{2} + r\right)^2 \tag{E5}$$

and $H_{\mathrm{eff}}$ is Hermitian. The stationary state Schrödinger equation for option price under $H_{\mathrm{eff}}$ is

$$\begin{cases} -\dfrac{\sigma^2}{2}\dfrac{\mathrm{d}^2\phi}{\mathrm{d}x^2} + \gamma\phi = E\phi, & x < B, \\ -\dfrac{\sigma^2}{2}\dfrac{\mathrm{d}^2\phi}{\mathrm{d}x^2} + (\gamma + V_0)\phi = E\phi, & x \geq B. \end{cases} \tag{E6}$$

where $\phi$ is the stationary state wave function of the option, and $E$ is corresponding to the option energy under the potential (E1). The Schrödinger equation above could be simplified into

$$\begin{cases} \dfrac{\mathrm{d}^2\phi}{\mathrm{d}x^2} + p_1^2\phi = 0, & x < B, \\ \dfrac{\mathrm{d}^2\phi}{\mathrm{d}x^2} - p_2^2\phi = 0, & x \geq B. \end{cases} \tag{E7}$$

where

$$p_1^2 = \frac{2(E - \gamma)}{\sigma^2}, \ \ p_2^2 = \frac{2(V_0 + \gamma - E)}{\sigma^2} \tag{E8}$$

here the condition $E < V_0$ has been considered, and the range of $p_1^2$ and $p_2^2$ is

$$\begin{aligned} p_2^2 &> \frac{2\gamma}{\sigma^2} \\ p_1^2 &= \frac{2V_0}{\sigma^2} - p_2^2 < \frac{2(V_0 - \gamma)}{\sigma^2} \end{aligned} \tag{E9}$$

The general solution for (E7) is

$$\phi(x) = \begin{cases} e^{ip_1(x-B)} + A_1 e^{-ip_1(x-B)}, & x < B, \\ A_2 e^{-p_2(x-B)}, & x \le B. \end{cases} \tag{E10}$$

where

$$A_1 = \frac{p_1 - ip_2}{p_1 + ip_2}, \quad A_2 = \frac{2p_1}{p_1 + ip_2} \tag{E11}$$

here the boundary condition at $x = B$ is used. Now we calculate the price of up-and-out proportional step call. Let $\tau_1$ indicates the occupation time below the barrier $B$, and $\tau_2$ is the occupation time above the barrier $B$. The pricing kernel is

$$\begin{aligned} p_{\text{PSO}}(x, x'; \tau) &= \langle x | e^{-\tau_1 H_1 - \tau_2 H_2} | x' \rangle \\ &= e^{\alpha(x-x')} \int \frac{dp}{2\pi} \langle x | e^{-\tau_1 H_{\text{eff1}}} | p \rangle \langle p | e^{-\tau_2 H_{\text{eff2}}} | x' \rangle \end{aligned} \tag{E12}$$

and the option price is

$$C(x; \tau) = \int_{\ln K}^{+\infty} dx' \, p_{\text{PSO}}(x, x'; \tau)(e^{x'} - K) \tag{E13}$$

where $K$ is the exercise price, and

$$\begin{aligned} H_1 &= -\frac{\sigma^2}{2} \frac{\partial^2}{\partial x^2} + \left(\frac{1}{2}\sigma^2 - r\right) \frac{\partial}{\partial x} + r \\ H_2 &= -\frac{\sigma^2}{2} \frac{\partial^2}{\partial x^2} + \left(\frac{1}{2}\sigma^2 - r\right) \frac{\partial}{\partial x} + r + V_0 \\ H_{\text{eff1}} &= -\frac{\sigma^2}{2} \frac{\partial^2}{\partial x^2} + \gamma \\ H_{\text{eff2}} &= -\frac{\sigma^2}{2} \frac{\partial^2}{\partial x^2} + \gamma + V_0 \end{aligned} \tag{E14}$$

We set the exercise price $K < e^B$. For $x < B$ and $x > B$, the wave functions are different, and the integral for (E13) should be split into four cases

- $x < B$, and $\ln K < x' < B$ the pricing kernel

$$\begin{aligned} & p_{\text{PSO1}}(x, x'; \tau) \\ &= e^{-\tau\gamma} e^{\alpha(x-x')} \int_0^{\frac{\sqrt{2(V_0-\gamma)}}{\sigma}} \frac{dp_1}{2\pi} e^{-\frac{1}{2}\tau\sigma^2 p_1^2} \left( e^{ip_1(x-B)} + A_1 e^{-ip_1(x-B)} \right) \\ & \times \left( e^{-ip_1(x'-B)} + A_1^* e^{ip_1(x'-B)} \right) \\ &= e^{-\tau\gamma} e^{\alpha(x-x')} \int_0^{\frac{\sqrt{2(V_0-\gamma)}}{\sigma}} \frac{dp_1}{2\pi} e^{-\frac{1}{2}\tau\sigma^2 p_1^2} \Bigg[ 2\cos p_1(x - x') \\ & + \frac{\sigma^2}{V_0}(p_1^2 - p_2^2) \cos p_1(x + x' - 2B) \\ & - \frac{2\sigma^2}{V_0} p_1 p_2 \sin p_1(x + x' - 2B) \Bigg] \end{aligned} \tag{E15}$$

- $x < B$, and $x' > B$

  the pricing kernel

  $p_{\mathrm{PSO2}}(x, x'; \tau)$

  $$
  = e^{-\tau\gamma} e^{\alpha(x-x')} \int_0^{\frac{\sqrt{2(V_0-\gamma)}}{\sigma}} \frac{\mathrm{d}p_1}{2\pi} e^{-\frac{1}{2}\tau\sigma^2 p_1^2} \left( e^{ip_1(x-B)} + A_1 e^{-ip_1(x-B)} \right) A_2^* e^{-p_2(x'-B)}
  $$

  $$
  = 2e^{-\tau\gamma} e^{\alpha(x-x')} \int_0^{\frac{\sqrt{2(V_0-\gamma)}}{\sigma}} \frac{\mathrm{d}p_1}{2\pi} e^{-\frac{1}{2}\tau\sigma^2 p_1^2} \left[ \frac{\sigma^2}{V_0} p_1^2 \cos p_1(x-B) \right.
  $$

  $$
  \left. - \frac{\sigma^2}{V_0} p_1 p_2 \sin p_1(x-B) \right]
  $$

  $$
  \times e^{-p_2(x'-B)} \tag{E16}
  $$

  the option price for $x < B$ is

  $$
  C(x; \tau)|_{x<B} = C_1(x; \tau) + C_2(x; \tau) \tag{E17}
  $$

  where

  $$
  \begin{aligned}
  C_1(x, \tau) &= \int_{\ln K}^{B} p_{\mathrm{PSO1}}(x, x'; \tau)(e^{x'} - K) \\
  C_2(x, \tau) &= \int_{B}^{+\infty} p_{\mathrm{PSO2}}(x, x'; \tau)(e^{x'} - K)
  \end{aligned} \tag{E18}
  $$

- $x > B$, and $\ln K < x' < B$

  the pricing kernel

  $p_{\mathrm{PSO3}}(x, x'; \tau)$

  $$
  = e^{-\tau\gamma} e^{\alpha(x-x')} \int_0^{\frac{\sqrt{2(V_0-\gamma)}}{\sigma}} \frac{\mathrm{d}p_1}{2\pi} e^{-\frac{1}{2}\tau\sigma^2 p_1^2} A_2 e^{-p_2(x-B)} \left( e^{-ip_1(x'-B)} + A_1^* e^{ip_1(x'-B)} \right)
  $$

  $$
  = 2e^{-\tau\gamma} e^{\alpha(x-x')} \int_0^{\frac{\sqrt{2(V_0-\gamma)}}{\sigma}} \frac{\mathrm{d}p_1}{2\pi} e^{-\frac{1}{2}\tau\sigma^2 p_1^2} e^{-p_2(x-B)} \left[ \frac{\sigma^2}{V_0} p_1^2 \cos p_1(x'-B) \right.
  $$

  $$
  \left. - \frac{\sigma^2}{V_0} p_1 p_2 \sin p_1(x'-B) \right] \tag{E19}
  $$

- $x > B$, and $x' > B$

  $$
  p_{\mathrm{PSO4}}(x, x'; \tau)
  $$

  $$
  = e^{-\tau\gamma} e^{\alpha(x-x')} \int_0^{\frac{\sqrt{2(V_0-\gamma)}}{\sigma}} \frac{\mathrm{d}p_1}{2\pi} e^{-\frac{1}{2}\tau\sigma^2 p_1^2} A_2 e^{-p_2(x-B)} A_2^* e^{-p_2(x'-B)}
  $$

  $$
  = 2e^{-\tau\gamma} e^{\alpha(x-x')} \int_0^{\frac{\sqrt{2(V_0-\gamma)}}{\sigma}} \frac{\mathrm{d}p_1}{2\pi} e^{-\frac{1}{2}\tau\sigma^2 p_1^2} \frac{\sigma^2}{V_0} p_1^2 e^{-p_2(x+x'-2B)} \tag{E20}
  $$

the option price for $x > B$ is

$$
C(x; \tau)|_{x>B} = C_3(x; \tau) + C_4(x; \tau) \tag{E21}
$$

where

$$C_3(x, \tau) = \int_{\ln K}^{B} p_{\text{PSO3}}(x, x'; \tau)(e^{x'} - K)$$
$$C_4(x, \tau) = \int_{B}^{+\infty} p_{\text{PSO4}}(x, x'; \tau)(e^{x'} - K) \tag{E22}$$

## Appendix F: Hamiltonian Approach to Proportional Double-Barrier Step Option Pricing

The price changing of a proportional double-barrier step (PDBS) option could be analogous to a particle moving in a symmetric square potential well with the potential

$$V(x) = \begin{cases} 0, & a < x < b, \\ V_0, & x < a, \ x > b. \end{cases} \tag{F1}$$

for $x < a$ or $x > b$, the wave function decays with the increasing distances from the well, which is similar to an option touches a barrier and knocks out gradually. The Hamiltonian for a double-barrier step option is [7]

$$H_{\text{PDBS}} = -\frac{\sigma^2}{2}\frac{\partial^2}{\partial x^2} + \left(\frac{1}{2}\sigma^2 - r\right)\frac{\partial}{\partial x} + r + V(x) \tag{F2}$$

which is a non-Hermitian Hamiltonian. Considering the following transformation

$$H_{\text{PDBS}} = e^{\alpha x} H_{\text{eff}} e^{-\alpha x} = e^{\alpha x}\left(-\frac{\sigma^2}{2}\frac{\partial^2}{\partial x^2} + \gamma\right)e^{-\alpha x} + V(x) \tag{F3}$$

where

$$\alpha = \frac{1}{\sigma^2}\left(\frac{\sigma^2}{2} - r\right), \gamma = \frac{1}{2\sigma^2}\left(\frac{\sigma^2}{2} + r\right)^2 \tag{F4}$$

and $H_{\text{eff}}$ is a Hermitian Hamiltonian which is considered as the symmetric square potential well Hamiltonian. The stationary state Schrödinger equation for option price is

$$\begin{cases} -\dfrac{\sigma^2}{2}\dfrac{\mathrm{d}^2\phi}{\mathrm{d}x^2} + \gamma\phi = E\phi, & a < x < b, \\ -\dfrac{\sigma^2}{2}\dfrac{\mathrm{d}^2\phi}{\mathrm{d}x^2} + (\gamma + V_0)\phi = E\phi, & x < a, \ x > b. \end{cases} \tag{F5}$$

where $\phi$ is the option price wave function, $E$ is corresponding to bound state energy levels in the potential well. (F5) could be simplified into

$$\begin{cases} \dfrac{\mathrm{d}^2\phi}{\mathrm{d}x^2} + k_1^2\phi = 0, & a < x < b, \\ \dfrac{\mathrm{d}^2\phi}{\mathrm{d}x^2} - k_2^2\phi = 0, & x < a, \ x > b. \end{cases} \tag{F6}$$

where

$$k_1^2 = \frac{2(E - \gamma)}{\sigma^2}, \quad k_2^2 = \frac{2(V_0 + \gamma - E)}{\sigma^2} \tag{F7}$$

The general solution for (F6) is

$$\phi(x) = \begin{cases} A_3\, e^{k_2(x - \frac{b+a}{2})}, & x \le a, \\ A_1 \sin(k_1 x + \delta), & a < x \le b, \\ A_2\, e^{-k_2(x - \frac{b+a}{2})}, & x > b. \end{cases} \qquad \text{(F8)}$$

Considering the continuity for both wave function and its derivative at $x = a$ and $x = b$, we have

$$\delta = \frac{\ell\pi}{2} - k_1 \frac{b+a}{2}, \ \ \ell = 0, 1, 2, ... \qquad \text{(F9)}$$

According to different $\ell s$ in (F9), (F8) could be split into two parts

$$\phi_1(x) = \begin{cases} -A_2 e^{k_2(x - \frac{b+a}{2})}, & x \le a, \\ A_1 \sin k_1\left(x - \frac{b+a}{2}\right), & a \le x \le b, \quad for\ \ell = 0, 2, 4, ... \\ A_2 e^{-k_2(x - \frac{b+a}{2})}, & x > b. \end{cases} \qquad \text{(F10)}$$

and

$$\phi_2(x) = \begin{cases} A_2 e^{k_2(x - \frac{b+a}{2})}, & x \le a, \\ A_1 \cos k_1\left(x - \frac{b+a}{2}\right), & a \le x \le b, \quad for\ \ell = 1, 3, 5, ... \\ A_2 e^{-k_2(x - \frac{b+a}{2})}, & x > b. \end{cases} \qquad \text{(F11)}$$

where

$$A_1 = \sqrt{\frac{2k_2}{k_2(b-a) + 2}}, \ \ A_2 = A_1 \sin\left(k_1 \frac{b-a}{2}\right) e^{k_2 \frac{b-a}{2}} \qquad \text{(F12)}$$

here the normalization condition has been used. Considering boundary conditions for (F10) and (F11) at $x = b$ respectively, we have

$$\cot k_1 \frac{b-a}{2} = -\frac{k_2}{k_1} \qquad \text{(F13)}$$

$$\tan k_1 \frac{b-a}{2} = \frac{k_2}{k_1} \qquad \text{(F14)}$$

let

$$\beta = \sqrt{k_1^2 + k_2^2} = \frac{\sqrt{2V_0}}{\sigma} \qquad \text{(F15)}$$

(F13) and (F14) could be combined into

$$k_{1n} \frac{b-a}{2} = \frac{n\pi}{2} - \arcsin\frac{k_{1n}}{\beta}, \ \ n = 1, 2, 3, ... \qquad \text{(F16)}$$

(F16) is the energy level equation. In general, there is no analytical solution for energy eigenvalues, yet numerical results for $k_{1n}$ could be obtained by Mathematica. In Table 1, we show $k_{1n}$ for different $n$s at $V_0 = 55.7859, 26.3401,$ and $12.8233$, which are corresponding to $d = 0.8, 0.9$ and $0.95$ in [12], respectively. The range of $k_{1n}$ could be obtained from (F7)

$$k_{1n} < \sqrt{\frac{2(V_0 - \gamma)}{\sigma^2}} \qquad \text{(F17)}$$

**Table 1** $k_{1n}$ for different $n$s at $V_0 = 55.7859, 26.3401$

| $V_0$ | $n$ | $k_{1n}$ | $n$ | $k_{1n}$ |
|---|---|---|---|---|
| 55.7859 | 1 | 7.38515 | 4 | 28.8819 |
|  | 2 | 14.7215 | 5 | 34.4789 |
|  | 3 | 21.9384 | 6 | 42.1887 |
| 26.3401 | 1 | 6.94708 | 3 | 20.1958 |
|  | 2 | 13.7663 | 4 | 25.0978 |
| 12.8233 | 1 | 6.41782 | 2 | 12.527 |

Parameters: $\sigma = 0.3, r = 0.05$

which restricts the number of energy levels is 5, 3 and 2 for $V_0 = 55.7859, 26.3401$, and 12.8233, respectively.

Now we calculate the pricing kernel of proportional double-barrier step option. a and b in (F1) could be considered as the lower and the upper barriers of the option. Let $\tau_1$ indicates the occupation time between the lower barrier a and the upper barrier b, and $\tau_2$ is the occupation time below the lower barrier a and above the upper barrier b. The pricing kernel is

$$
\begin{aligned}
P_{\text{PDBS}}(x, x'; \tau) &= \langle x | e^{-\tau_1 H_1 - \tau_2 H_2} | x' \rangle \\
&= \int_{-\infty}^{+\infty} \mathrm{d}x'' \langle x | e^{-\tau_1 H_1} | x'' \rangle \langle x'' | e^{-\tau_2 H_2} | x' \rangle \\
&= e^{-\tau \gamma} \int_{-\infty}^{+\infty} \mathrm{d}x'' e^{\alpha(x - x')} \sum_n \sum_{n'} e^{-\tau_1 E_{1n} - \tau_2 E_{2n'}} \phi_n(x) \phi_n(x'') \phi_{n'}(x'') \phi_{n'}(x')
\end{aligned}
$$
(F18)

where $S_0 = e^x$ is the initial price of the underlying asset, $S_T = e^{x'}$ is the final price of the underlying asset. Using the orthonormalization condition

$$
\int_{-\infty}^{+\infty} \mathrm{d}x'' \, \phi_n(x'') \phi_{n'}(x'') = \delta_{nn'} = \begin{cases} 0, & n \neq n', \\ 1, & n = n'. \end{cases}
$$
(F19)

the pricing kernel (F18) is simplified into

$$
P_{\text{PDBS}}(x, x'; \tau) = e^{-\tau \gamma} e^{\alpha(x - x')} \sum_n e^{-\frac{1}{2}\tau \sigma^2 k_{1n}^2} \phi_n(x) \phi_n(x')
$$
(F20)

where (F7) is used, and the proportional double-barrier call price is

$$
C_{\text{PDBS}}(x; \tau) = \int_{\ln K}^{+\infty} \mathrm{d}x' p_{\text{PDBS}}(x, x'; \tau)(e^{x'} - K)
$$
(F21)

where

$$
\begin{aligned}
H_1 &= e^{\alpha x}\left(-\frac{\sigma^2}{2}\frac{\partial^2}{\partial x^2} + \gamma\right) e^{-\alpha x} \\
H_2 &= e^{\alpha x}\left(-\frac{\sigma^2}{2}\frac{\partial^2}{\partial x^2} + \gamma\right) e^{-\alpha x} + V_0
\end{aligned}
$$
(F22)

$\phi_n(x)$ is the energy eigenstate in coordinate representation, $K$ is the exercise price, and $\tau = \tau_1 + \tau_2$ is the time to expiration. Setting $\ln K \in (a, b)$, combining with the expressions of $\phi(x)$ in (F10) and (F11), the option price for $a < x < b$ could be denoted as

$$C(x,\tau)\big|_{a<x<b} = \int_{\ln K}^{b} dx' P_{\text{PDBS}}(x,x';\tau)\big|_{a<x<b,\ln K<x'<b}(e^{x'} - K)$$
$$+ \int_{b}^{+\infty} P_{\text{PDBS}}(x,x';\tau)\big|_{a<x<b,x'>b}(e^{x'} - K) \qquad \text{(F23)}$$

where

$$P_{\text{PDBS}}(x,x';\tau)\big|_{a<x<b,\ln K<x'<b}$$
$$= e^{-\tau\gamma}e^{\alpha(x-x')}(e^{x'} - K)$$
$$\times \left( \sum_{n=2,4,6...} e^{-\frac{1}{2}\tau\sigma^2 k_{1n}^2} A_1^2 \sin\left[k_{1n}\left(x - \frac{b+a}{2}\right)\right]\sin\left[k_{1n}\left(x' - \frac{b+a}{2}\right)\right]\right.$$
$$\left.+ \sum_{n=1,3,5...} e^{-\frac{1}{2}\tau\sigma^2 k_{1n}^2} A_1^2 \cos\left[k_{1n}\left(x - \frac{b+a}{2}\right)\right]\cos\left[k_{1n}\left(x' - \frac{b+a}{2}\right)\right]\right) \quad \text{(F24)}$$

$$P_{\text{PDBS}}(x,x';\tau)\big|_{a<x<b,x'>b}$$
$$= e^{-\tau\gamma}e^{\alpha(x-x')}(e^{x'} - K)$$
$$\times \left( \sum_{n=2,4,6...} e^{-\frac{1}{2}\tau\sigma^2 k_{1n}^2} A_1 A_2 \sin\left[k_{1n}\left(x - \frac{b+a}{2}\right)\right]e^{-k_{2n}(x'-\frac{b+a}{2})}\right.$$
$$\left.+ \sum_{n=1,3,5...} e^{-\frac{1}{2}\tau\sigma^2 k_{1n}^2} A_1 A_2 \cos\left[k_{1n}\left(x - \frac{b+a}{2}\right)\right]e^{-k_{2n}(x'-\frac{b+a}{2})}\right) \qquad \text{(F25)}$$

Similarly, the option price for $x > a$ is

$$C(x,\tau)\big|_{x>a} = \int_{\ln K}^{b} dx' P_{\text{PDBS}}(x,x';\tau)\big|_{x>a,\ln K<x'<b}(e^{x'} - K)$$
$$+ \int_{b}^{+\infty} P_{\text{PDBS}}(x,x';\tau)\big|_{x>a,x'>b}(e^{x'} - K) \qquad \text{(F26)}$$

where

$$P_{\text{PDBS}}(x,x';\tau)\big|_{x>a,\ln K<x'<b}$$
$$= e^{-\tau\gamma}e^{\alpha(x-x')}(e^{x'} - K)$$
$$\times \left( \sum_{n=1,3,5...} e^{-\frac{1}{2}\tau\sigma^2 k_{1n}^2} A_2 A_1 e^{-k_{2n}(x-\frac{b+a}{2})}\sin\left[k_{1n}\left(x' - \frac{b+a}{2}\right)\right]\right.$$
$$\left.+ \sum_{n=2,4,6...} e^{-\frac{1}{2}\tau\sigma^2 k_{1n}^2} A_2 A_1 e^{-k_{2n}(x-\frac{b+a}{2})}\cos\left[k_{1n}\left(x' - \frac{b+a}{2}\right)\right]\right) \qquad \text{(F27)}$$

$$P_{\text{PDBS}}(x,x';\tau)\big|_{x>a,x'>b}$$
$$= e^{-\tau\gamma}e^{\alpha(x-x')}(e^{x'} - K)$$
$$\times \left( \sum_{n=1,3,5...} e^{-\frac{1}{2}\tau\sigma^2 k_{1n}^2} A_2^2 e^{-k_{2n}(x-\frac{b+a}{2})}e^{-k_{2n}(x'-\frac{b+a}{2})}\right.$$
$$\left.+ \sum_{n=2,4,6...} e^{-\frac{1}{2}\tau\sigma^2 k_{1n}^2} A_2^2 e^{-k_{2n}(x-\frac{b+a}{2})}e^{-k_{2n}(x'-\frac{b+a}{2})}\right) \qquad \text{(F28)}$$